\documentclass[runningheads]{llncs}
\usepackage{graphicx}
\usepackage{hyperref}
\usepackage{url}
\usepackage{pgfplots}
\pgfplotsset{width=12cm,compat=1.9}
\usepackage{listings}

\begin{document}
\title{A Preliminary Analysis on the Code Generation Capabilities of GPT-3.5 and Bard AI Models for Java Functions}

\author{
Giuseppe Destefanis$^{1,\star}$, Silvia Bartolucci$^2$ and Marco Ortu$^3$\\
$^1$\textit{Dept. of Computer Science, Brunel University London, UK}\\
$^2$\textit{Dept. of Computer Science, University College London, UK}\\
$^3$\textit{Department of Business and Economics Sciences, University of Cagliari, IT}\\
$ \star$ {\small Corresponding author: \texttt{giuseppe.destefanis@brunel.ac.uk}}\\\
}

\maketitle              % typeset the header of the contribution
\begin{abstract}
This paper evaluates the capability of two state-of-the-art artificial intelligence (AI) models, GPT-3.5 and Bard, in generating Java code given a function description. We sourced the descriptions from CodingBat.com, a popular online platform that provides practice problems to learn programming. We compared the Java code generated by both models based on correctness, verified through the platform's own test cases. The results indicate clear differences in the capabilities of the two models. GPT-3.5 demonstrated superior performance, generating correct code for approximately 90.6\% of the function descriptions, whereas Bard produced correct code for 53.1\% of the functions. While both models exhibited strengths and weaknesses, these findings suggest potential avenues for the development and refinement of more advanced AI-assisted code generation tools. The study underlines the potential of AI in automating and supporting aspects of software development, although further research is required to fully realize this potential.

% \keywords{ \and Second keyword \and Another keyword.}
\end{abstract}

\section{Important Notice}
It is crucial to highlight that the experiments conducted in this study took place between the 10th and the 15th of May 2023. The AI systems analyzed—GPT-3.5 and Bard—are continually learning and evolving systems. As they are fed more data and undergo further training, their performance and capabilities can significantly change over time. Consequently, the results and comparisons drawn in this study are, by their nature, temporally bound and may not hold true in the future. Furthermore, it should be noted that Bard is an experimental tool, as explicitly stated by the provider. Therefore, its performance, behavior, and even availability may fluctuate. As such, readers are advised to interpret the results and conclusions of this paper in light of these considerations.

\newpage

\section{Introduction}
\label{sec:introduction}
Artificial Intelligence (AI) has made significant strides in recent years, particularly in the area of natural language understanding and generation. 
With the widespread proliferation of large language models capable of performing intricate tasks, research efforts have been undertaken to benchmark and compare the performance of these models \cite{survey}.  Our analysis focuses on two specific AI models, GPT-3.5 and Bard, investigating their capabilities to generate Java code based on function descriptions obtained from CodingBat.com.

In previous research, the mathematical reasoning capabilities of Generative Pre-trained Tranformers (GPTs) were assessed against state-of-the-art large language models (LLMs) \cite{GPTmaths}. The system was presented with a set of mathematics questions taken, and the output was rated manually to assess their performance. In \cite{Bardneuro} GPT-3.5, 4 and Bard's performances in answering medical questions from a neurosurgery question bank were compared, showing that GPT models systematically outperformed Bard.
GPT-3.5 and 4 were also shown to be able to pass complex tests, such as the Bar Exam in U.S., with GPT-4 consistently surpassing its LLMs predecessors as well as previous recorded students' performance \cite{Katz}.  GPT-3 models were also tested for code vulnerability and bug detection, showing that their capabilities were limited \cite{codebug}. Similarly, GPT-3.5 and GPT-4 abilities were tested against other LLMs on coding tasks \cite{code}, performing translation in different languages \cite{Translation}, and stock return forecasting \cite{pricepred}.

\subsection{GPT-3.5}

The \href{https://platform.openai.com/docs/models/gpt-3}{GPT-3.5} model is a variant of the transformative Generative Pretrained Transformer 3 (GPT-3) model. This language prediction model utilizes a transformer architecture to understand and generate human-like text based on given prompts. GPT-3.5 has 175 billion parameters and has been trained on a diverse range of internet text. Its large scale allows it to generate coherent and contextually relevant sentences over long passages. However, it is important to note that GPT-3.5 does not have any specific knowledge or understanding of the content; it generates responses based on patterns learned during its training.

\subsection{Bard}

\href{https://bard.google.com}{Bard} is another advanced AI model known for its exceptional natural language generation capabilities. While details of its architecture and training data are proprietary, Bard has demonstrated strong performance in various natural language understanding and generation tasks. It  is particularly noted for its ability to generate creative and contextually nuanced text. Similar to GPT-3.5, Bard does not have a conscious understanding of the content it generates but instead identifies and follows patterns in the data it was trained on.

\subsection{CodingBat.com}

\href{https://codingbat.com}{CodingBat} is an online platform that offers coding practice problems to help students learn and master various programming concepts. The platform primarily focuses on Java and Python problems. The problems on CodingBat.com cover a wide range of difficulty levels and programming concepts, making it an ideal resource for learners of all levels. The solutions to problems are checked and run in real-time, providing immediate feedback to the learners.

This analysis assesses the capabilities of the GPT-3.5 and Bard AI models by feeding them with function descriptions from CodingBat.com. The generated Java code from both models is subsequently evaluated exclusively in terms of correctness, assessed using the real-time code checking system of CodingBat.com. The objective of this research is to analyse the respective strengths and weaknesses of GPT-3.5 and Bard in terms of generating correct Java code. The insights derived from this preliminary study could be helpful in the creation of more refined, AI-assisted code generation tools in the future.

\section{Methodology}

\subsection{Data Collection}

For this study, we selected Java function descriptions from five distinct sections of CodingBat.com, each section representing a different level of problem complexity and topic. This selection aimed at ensuring a diverse range of problems that would challenge the code generation capabilities of both the GPT-3.5 and Bard AI models.

The selected sections were as follows:

\begin{itemize}
\item {\bf Warmup}: This section contains basic Java functions that are designed as warm up on programming skills. Despite their simplicity, these problems provide a good starting point for testing the AI models' grasp of fundamental programming concepts.
\item {\bf String-3}: This section presents harder string problems that usually involve the use of two loops. The challenges in this section test the AI models' ability to handle more complex string manipulation tasks and control structures.
\item {\bf Array-3}: This section contains harder array problems, also involving two loops and more complex logic. These problems allow us to evaluate how well the AI models can generate code for more advanced array manipulation tasks and complex logical operations.
\item {\bf Functional-2}: This section includes functional filtering and mapping operations on lists using Java lambdas. These tasks test the AI models' capability to generate code using functional programming concepts in Java, which differ significantly from imperative programming tasks.
\item {\bf Recursion-2}: This section contains harder recursion problems. These problems challenge the AI models to generate code that implements recursive solutions, which are significantly different in structure from iterative solutions and can be particularly challenging to generate.
\end{itemize}

From each of these five sections, we collected all the Java function descriptions, resulting in a total of $64$ function descriptions for our study. Each of these descriptions was provided as a prompt to both the GPT-3.5 and Bard AI models to generate the corresponding Java code. The generated code was subsequently evaluated based on its correctness, with the results provided by the real-time code checking system of CodingBat.com.

\section{Result}

In this analysis, we evaluated the Java code generated by GPT-3.5 (ChatGPT) and Bard AI models using function descriptions from five distinct sections of CodingBat.com. Each piece of code was evaluated for correctness, with `1' denoting correct code and `0' representing incorrect code. 

The results for the category {\bf WarmUp} are summarized in Table \ref{table:1}, as shown below:

\begin{table}[h]
\centering
\begin{tabular}{lcc}
\hline
\textbf{Function} & \textbf{GPT-3.5} & \textbf{Bard} \\
\hline
sleepIn & 1 & 1 \\
diff21 & 1 & 1 \\
parrotTrouble & 1 & 1 \\
makes10 & 1 & 1 \\
nearHundred & 1 & 1 \\
missingChar & 1 & 1 \\
frontBack & 1 & 0 \\
front3 & 1 & 0 \\
backaround & 1 & 1 \\
or35 & 1 & 1 \\
front22 & 1 & 0 \\
startHi & 1 & 1 \\
icyHot & 1 & 1 \\
in1020 & 1 & 1 \\
hasTeen & 1 & 1 \\
IoneTeen & 1 & 1 \\
delDel & 1 & 0 \\
mixStart & 1 & 1 \\
startOz & 1 & 0 \\
intMax & 1 & 1 \\
close10 & 1 & 1 \\
in3050 & 1 & 1 \\
max1020 & 1 & 1 \\
stringE & 1 & 1 \\
lastDigit & 1 & 1 \\
endUp & 1 & 1 \\
everyNth & 1 & 1 \\
\hline
\end{tabular}
\caption{Comparison of Java Code Correctness between GPT-3.5 and Bard for the category WarmUp.} 
\label{table:1}
\end{table}

From these results, it can be observed that GPT-3.5 was able to generate correct Java code for all the given function descriptions. On the other hand, Bard had a few instances where it produced incorrect code. Specifically, Bard was unable to generate correct code for the function descriptions \texttt{frontBack, front3, front22, delDel, and startOz}.

In the {\bf String-3} category, presented in Table \ref{table:2}, which involves more complex string manipulation problems, the GPT-3.5 model outperformed Bard. GPT-3.5 correctly generated code for all but one of the given function descriptions. In contrast, Bard only correctly generated code for two functions, \texttt{gHappy} and \texttt{sumDigit}. Notably, both models failed to generate correct code for the \texttt{notReplace} function, suggesting that this problem posed a significant challenge. 

\begin{table}[h]
\centering
\begin{tabular}{lcc}
\hline
\textbf{Function} & \textbf{GPT-3.5} & \textbf{Bard} \\
\hline
countYZ & 1 & 0 \\
withoutString & 1 & 0 \\
equalIsNot & 1 & 0 \\
gHappy & 1 & 1 \\
counTriple & 1 & 0 \\
sumDigit & 1 & 1 \\
sameEnds & 1 & 0 \\
mirrorEnds & 1 & 0 \\
maxBlock & 1 & 0 \\
sumNumber & 1 & 0 \\
notReplace & 0 & 0 \\
\hline
\end{tabular}
\caption{Comparison of Java Code Correctness for String-3 category between GPT-3.5 and Bard.}
\label{table:2}
\end{table}

In the {\bf Array-3} category, Table \ref{table:3}, which contains harder array problems requiring more complex logic and double loops, GPT-3.5 demonstrated stronger performance compared to Bard. GPT-3.5 correctly generated code for five out of nine function descriptions. On the other hand, Bard struggled with these more complex tasks and was unable to correctly generate code for any of the problems in this category. Both models, however, were unable to generate correct code for the function descriptions \texttt{fix34, fix45}, and \texttt{squareUp}.

\begin{table}[h]
\centering
\begin{tabular}{lcc}
\hline
\textbf{Function} & \textbf{GPT-3.5} & \textbf{Bard} \\
\hline
maxSpan & 1 & 0 \\
fix34 & 0 & 0 \\
fix45 & 0 & 0 \\
canBalance & 1 & 0 \\
linearIn & 1 & 0 \\
squareUp & 0 & 0 \\
seriesUp & 1 & 0 \\
maxMirror & 1 & 0 \\
countClumps & 1 & 0 \\
\hline
\end{tabular}
\caption{Comparison of Java Code Correctness for Array-3 category between GPT-3.5 and Bard.}
\label{table:3}
\end{table}

In the {\bf Functional-2} category, Table \ref{table:4}, which includes functional filtering and mapping operations on lists with lambdas, both GPT-3.5 and Bard demonstrated high levels of competence and managed to correctly generate code for all the function descriptions. These results suggest that both GPT-3.5 and Bard have robust capabilities when it comes to handling functional programming tasks.

\begin{table}[]
\centering
\begin{tabular}{lcc}
\hline
\textbf{Function} & \textbf{GPT-3.5} & \textbf{Bard} \\
\hline
noNeg & 1 & 1 \\
no9 & 1 & 1 \\
noTeen & 1 & 1 \\
noZ & 1 & 1 \\
noLong & 1 & 1 \\
no34 & 1 & 1 \\
noYY & 1 & 1 \\
two2 & 1 & 1 \\
square56 & 1 & 1 \\
\hline
\end{tabular}
\caption{Comparison of Java Code Correctness for Functional-2 category between GPT-3.5 and Bard.}
\label{table:4}
\end{table}

In the {\bf Recursion-2} category, Table \ref{table:5}, which involves harder recursion problems, GPT-3.5 demonstrated relative strength with correct code generated for five out of the eight function descriptions provided. Bard, on the other hand, only managed to correctly generate code for the function \texttt{groupSum}. Notably, both models struggled with the problems \texttt{groupSum5} and \texttt{split53}, suggesting a higher level of difficulty in these problems.

\begin{table}[h]
\centering
\begin{tabular}{lcc}
\hline
\textbf{Function} & \textbf{GPT-3.5} & \textbf{Bard} \\
\hline
groupSum & 1 & 1 \\
groupSum6 & 1 & 0 \\
groupNoAdj & 1 & 0 \\
groupSum5 & 0 & 0 \\
groupSumClump & 1 & 0 \\
splitArray & 1 & 0 \\
splitOdd10 & 1 & 0 \\
split53 & 0 & 0 \\
\hline
\end{tabular}
\caption{Comparison of Java Code Correctness for Recursion-2 category between GPT-3.5 and Bard.}
\label{table:5}
\end{table}

In summary, across the four categories of problems Warmup, String-3, Array-3, Recursion-2, GPT-3.5 consistently outperformed Bard.
For the category Functional-2, both tools generated correct code.

GPT-3.5 generated correct code for 90.6\% of the problems, compared to Bard's 53.1\%. These results suggest that GPT-3.5 has a stronger ability to generate correct Java code based on function descriptions from a diverse range of problem types and complexities. This analysis also highlights that both models have room for improvement, particularly in handling certain types of complex problems. 

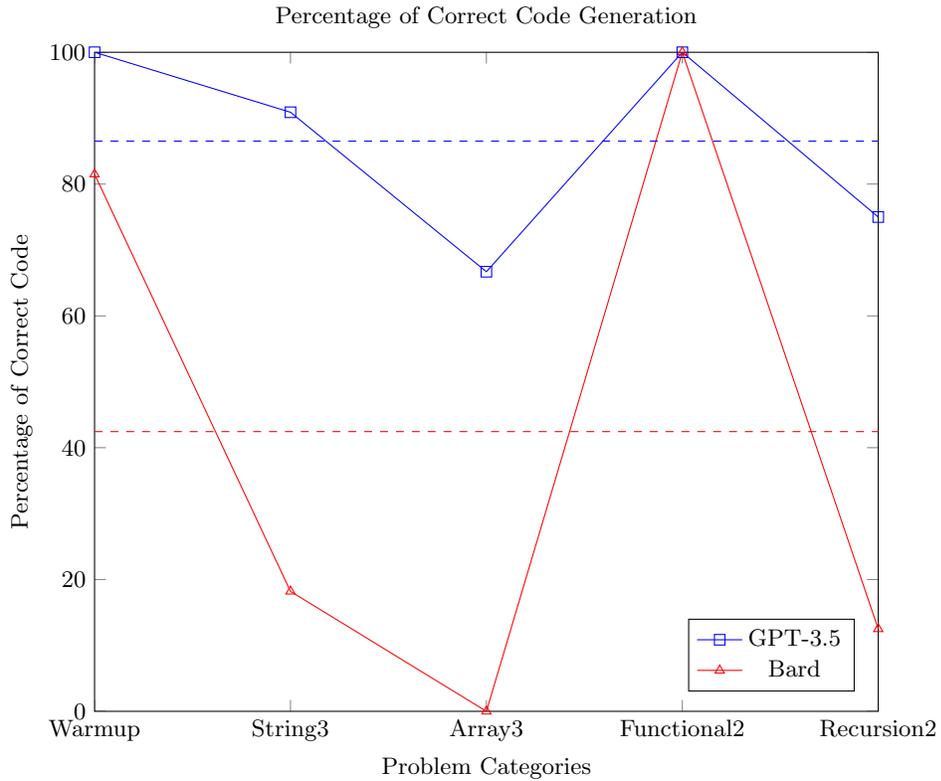
\begin{figure}
\centering
\begin{tikzpicture}
\begin{axis}[
    title={Percentage of Correct Code Generation},
    xlabel={Problem Categories},
    ylabel={Percentage of Correct Code},
    xmin=1, xmax=5,
    ymin=0, ymax=100,
    xtick={1,2,3,4,5},
    xticklabels={Warmup, String3, Array3, Functional2, Recursion2},
    ytick={0,20,40,60,80,100},
    legend pos=south east,
    ymajorgrids=false,
    grid style=dashed,
]

\addplot[
    color=blue,
    mark=square,
    ]
    coordinates {
    (1,100)(2,90.9)(3,66.7)(4,100)(5,75)
    };
    \addlegendentry{GPT-3.5}
    
\addplot[
    color=red,
    mark=triangle,
    ]
    coordinates {
    (1,81.5)(2,18.2)(3,0)(4,100)(5,12.5)
    };
    \addlegendentry{Bard}

\addplot[
    color=blue,
    mark=none,
    dashed,
]
coordinates {
    (1,86.52)(2,86.52)(3,86.52)(4,86.52)(5,86.52)
};

\addplot[
    color=red,
    mark=none,
    dashed,
]
coordinates {
    (1,42.44)(2,42.44)(3,42.44)(4,42.44)(5,42.44)
};

\end{axis}
\end{tikzpicture}
\label{fig:trends}
\caption{Comparison of the percentage of corrected code generated respectively by GPT-3.5 (ble line) and Bard (red line) in five different problem categories. Dashed lines represent the average performance of GPT-3.5 (blue dashed line) and Bard (red dashed line).}
\end{figure}

Figure \ref{fig:trends} illustrates the results. The plot provides a clear visual representation of the performance of GPT-3.5 and Bard across five problem categories, namely Warmup, String3, Array3, Functional2, and Recursion2.
For GPT-3.5, the model achieves near-perfect performance in the Warmup and Functional2 categories, demonstrating its effectiveness in handling relatively simple and functional problems. Its performance on String3 and Recursion2 problems is also good, with a slight dip observed in the Array3 category.
Contrarily, Bard's performance varies more significantly across categories. This model performs well in the Warmup and perfectly in the Functional2 categories but struggles with more complex categories such as String3, Array3, and Recursion2. The Array3 category proves particularly challenging for Bard, where it fails to generate any correct code.
The graph overall highlights the superior overall performance of GPT-3.5 across four categories. However, as stated above, the performance of both models suggests room for improvement, particularly in handling complex problem categories.

% \begin{tikzpicture}
% \begin{axis}[
%     title={Percentage of Correct Code Generation},
%     xlabel={Problem Categories},
%     ylabel={Percentage of Correct Code},
%     xmin=1, xmax=5,
%     ymin=0, ymax=100,
%     xtick={1,2,3,4,5},
%     xticklabels={Warmup, String3, Array3, Functional2, Recursion2},
%     ytick={0,20,40,60,80,100},
%     legend pos=south east,
%     ymajorgrids=false,
%     grid style=dashed,
% ]

% \addplot[
%     color=blue,
%     mark=square,
%     ]
%     coordinates {
%     (1,100)(2,90.9)(3,66.7)(4,100)(5,75)
%     };
%     \addlegendentry{GPT-3.5}
    
% \addplot[
%     color=red,
%     mark=triangle,
%     ]
%     coordinates {
%     (1,81.5)(2,18.2)(3,0)(4,88.9)(5,12.5)
%     };
%     \addlegendentry{Bard}

% \end{axis}
% \end{tikzpicture}

%The graph provides a visual illustration of the performance trends of the GPT-3.5 and Bard AI models across different task categories. 
Despite GPT-3.5 outperforming Bard across 4 out of 5 categories, it is evident that both models exhibit a similar performance pattern. This trend indicates a shared difficulty in tackling certain types of tasks, specifically those involving complex array operations and recursion. This shared challenge underscores the complexity and intricacy of these tasks even for sophisticated AI tools. The congruence in their performance trends suggests that these tasks tap into areas where both models, despite their distinct underlying architectures, face significant difficulties. Consequently, these areas represent critical points of attention for future enhancements in AI-assisted code generation.

We also performed a McNemar test \cite{mcnemar} to probe the difference between the two AI models, GPT-3.5 and Bard, in terms of their success rates on the same set of function descriptions. McNemar test is suited for this purpose as it is designed for paired nominal data, such as our case where each function description serves as a paired observation evaluated by both AI models. In our study, McNemar test was applied to a 2x2 contingency table, representing the success and failure rates of both AI models. The null hypothesis for McNemar test is that the marginal probabilities of success for the two AI models are equal.

The McNemar Chi-square statistic was found to be equal to 14.062 with 1 degree of freedom, and the associated $p$-value was 0.0001768. Given this p-value is substantially less than the conventional threshold for statistical significance (0.05), we reject the null hypothesis of equal success rates. This result indicates a statistically significant difference in the success rates of GPT-3.5 and Bard in generating correct Java code. In particular, GPT-3.5 demonstrated a higher success rate compared to Bard.

This outcome underlines the differing strengths of the AI models and encourages further investigation into the factors contributing to these differences.

\subsection{Examination of Unsuccessful Code Generation}

As an example, we report below the generated function \texttt{front3} of the Warmup section for which Bard provides an incorrect solution. 

\begin{figure}[ht!]
\centering
\includegraphics[width=0.9\textwidth]{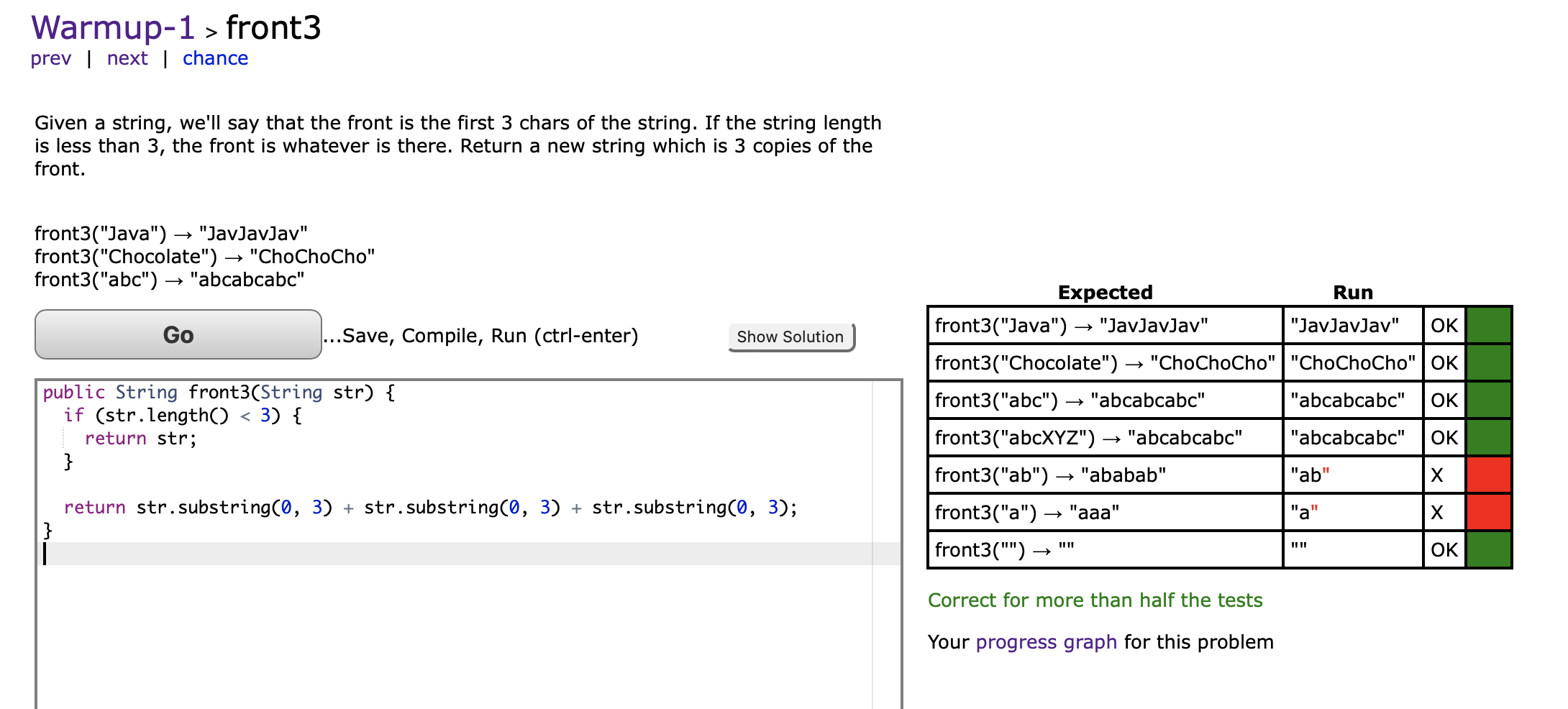}
\caption{Example of incorrect code generated by Bard.}
\label{fig:my_label}
\end{figure}

CodingBat asks the following:
\vspace{0.3cm}\\
\textit{Given a string, we'll say that the front is the first 3 chars of the string. If the string length is less than 3, the front is whatever is there. Return a new string which is 3 copies of the front.}

front3(``Java") ....``JavJavJav"

front3(``Chocolate") ...  ``ChoChoCho"

front3(``abc") ... ``abcabcabc" \vspace{0.3cm}\\
Bard prompt produced the following code:
\begin{lstlisting}
public String front3(String str) {
  if (str.length() < 3) {
    return str;
  }

  return str.substring(0, 3) + str.substring(0, 3) + str.substring(0, 3);
}
\end{lstlisting}

which produces the results on CodingBat illustrated in fig. \ref{fig:my_label}

GPT3.5 produced the following code:

\begin{lstlisting}
public String front3(String str) {
        if (str.length() < 3) {
            return str + str + str;
        } else {
            String front = str.substring(0, 3);
            return front + front + front;
        }
    }
\end{lstlisting}
which passed all tests performed by CodingbBat.

\section{Threats to Validity}

As with any empirical study, several threats to the validity of our results exist.

\subsection{Internal Validity}
The main internal threat comes from the fact that this study assumes that the AI models generate code that is correct based on the test cases provided by CodingBat.com. However, it is possible that the code could behave unexpectedly with other inputs, leading to potential overestimation of the models' capability. Additionally, the correctness of the generated code is determined only by the test cases from CodingBat.com. If these test cases are incomplete and do not cover all possible cases, this could also affect the results.

\subsection{External Validity}
In terms of external validity, while we have strived to ensure diversity in our function descriptions, they are nonetheless limited to the Java problems available on CodingBat.com. As such, these results may not generalize to other programming languages, problem domains, or more complex coding tasks. It is also worth noting that both AI models have been trained on diverse datasets, and their performance in this study might not reflect their performance in different contexts or tasks.

\subsection{Construct Validity}
This study assumes that the quality of code can be measured solely in terms of its correctness. However, other important aspects such as code efficiency, readability, maintainability, and adherence to best practices have not been considered. The absence of these metrics limits the comprehensive evaluation of the code quality generated by the AI models.

Despite these limitations, this study provides valuable insights into the capabilities of current AI models in generating Java code from function descriptions. Future work could focus on addressing these limitations by incorporating a wider range of problem types, considering other programming languages, and including more comprehensive measures of code quality.

\section{Future Work}

While this analysis provides a preliminary comparison of the code generation capabilities of GPT-3.5 and Bard, there are several directions for future research that could provide more comprehensive insights.
One critical aspect that needs further investigation is the complexity of the generated code. Evaluating the generated code not just on correctness but also on factors such as its efficiency, readability, and maintainability could provide a more nuanced understanding of the AI models' code generation capabilities. Advanced tools and techniques for evaluating code complexity should be utilized in future studies to assess code quality.
Furthermore, our study was restricted to Java programming language and the function descriptions available on CodingBat.com. Future studies should consider evaluating the AI models' performance with other programming languages and a wider range of problem types and complexities. This would allow for a more comprehensive evaluation of the code generation capabilities of these AI models.

It would also be beneficial to perform a longitudinal study, tracking the progress of these AI models over time. As these models continue to evolve and improve, it would be insightful to understand how their code generation capabilities change with each iteration.

\section{Conclusion}

This analysis sought to assess the capabilities of two AI models, GPT-3.5 and Bard, in generating Java code from function descriptions. Function descriptions were collected from five diverse categories of problems on CodingBat.com, and the AI models' output was evaluated based on correctness, as verified by the platform's own test cases.

Our findings reveal that GPT-3.5 outperformed Bard across four out of five categories, exhibiting overall a higher rate of correctly generated code. Specifically, GPT-3.5 achieved an overall correctness rate of approximately 90.6\%, compared to Bard's 53.1\%. This suggests that, at least within the context of this study, GPT-3.5 possesses a stronger ability to generate correct Java code based on diverse function descriptions.

However, it should also be noted that neither model was consistently correct in generating their output. Both models demonstrated shortcomings, particularly when faced with more complex problems such as those in the Recursion-2 category. This highlights  the need for continued development and refinement of AI code generation models.

Despite the inherent limitations and threats to validity, we believe that this study contributes valuable insights to the field of AI-assisted code generation. It provides a comparative evaluation of two popular AI models, and highlights areas of strength and potential improvement.

Looking forward, we envision that these findings will stimulate further research and advancement in the field. Specifically, we see potential for studies that explore a wider range of problem types, include other programming languages, and incorporate more comprehensive measures of code quality. Ultimately, this research will possibly contribute to the development of AI models that are capable of reliably assisting humans in diverse coding tasks, thereby enhancing productivity and fostering innovation in software development.

\bibliographystyle{unsrt}

\end{document}